\documentclass[twocolumn,aps,prl,superscriptaddress,floatfix,longbibliography]{revtex4-1}
\usepackage{graphicx}
\usepackage{amsmath}
\usepackage[colorlinks,linkcolor=blue,citecolor=blue,urlcolor=blue]{hyperref}

\begin{document}

% ======= Start with single column mode =======
\onecolumngrid

\begin{center}
    \textbf{\Large Same-group element replacement enhances superconductivity in clathrate-like YH$_4$}\\[10pt]
    
    Xuejie Li,\textsuperscript{1}
    Yuzhou Hao,\textsuperscript{1}
    Yujie Liu,\textsuperscript{1}
    Xiaoying Wang,\textsuperscript{1}
    Turab Lookman,\textsuperscript{1,2}
    Jun Sun,\textsuperscript{1}
    Xiangdong Ding,\textsuperscript{1}
    Zhibin Gao\textsuperscript{1,*}\\[6pt]

    \textsuperscript{1}\textit{State Key Laboratory for Mechanical Behavior of Materials, School of Materials Science and Engineering, Xi’an Jiaotong University, Xi’an 710049, China} \\
    \textsuperscript{2}\textit{AiMaterials Research LLC, Santa Fe, New Mexico 87501, USA} \\[6pt]
    \textsuperscript{*}Corresponding author: \texttt{zhibin.gao@xjtu.edu.cn}\\[12pt]

    (Dated: \today)
\end{center}

\vspace{12pt}

%---------------------------------------------------------------------
\begin{abstract}

%H$_3$S, LaH$_{10}$, and H-based compounds have drawn lots of interest since the high-temperature superconducting properties. Nevertheless, extremely high pressure may limit their further applications. Here, we adopt YH$_4$ as a base material and choose Scandium (Sc), Lanthanum (La), and Zirconium (Zr) as a substitution. We find YH$_4$ is stable at 120 GPa with a $T_c$ ranging from 84 K to 95 K. By substituting half of Y element, the $T_c$ has an increase to 104.36 $\sim$ 143 K for (Y,Sc)H$_4$ at 100 GPa, while has a decrease to 82.49 K $\sim$ 128 K for (Y,La)H$_4$ at 120 GPa. Moreover, $T_c$ is further suppressed to 52.61 K $\sim$ 100 K at 200 GPa for (Y,Zr)H$_4$, which is not as good as the finding for the YH$_4$. In particular, the good superconducting properties of (Y,Sc)H$_4$ is derived from the absence of gap between the optical and acoustic phonon branches, and shorter bond lengths relative to pressure. In addition, the superconductivity of (Y, Sc, La, Zr)H$_4$ is attributed to the crucial contribution of the lowest optical phonons. We find the substitution of same group metal elements may be an effective way to increase $T_c$ for hydride materials.

H$_3$S, LaH$_{10}$, and hydrogen-based compounds have garnered significant interest due to their high-temperature superconducting properties. However, the requirement for extremely high pressures limits their practical applications. In this study, YH$_4$ is adopted as a base material, with partial substitution of Yttrium (Y) by Scandium (Sc), Lanthanum (La), and Zirconium (Zr). Pure YH$_4$, stable at 120 GPa, exhibits a critical temperature ($T_c$) of 84-95 K. Substituting half of the Y atoms increases $T_c$ to 124.43 K for (Y,Sc)H$_4$ at 100 GPa but reduces it to 101.24 K for (Y,La)H$_4$ at 120 GPa. 
In contrast, (Y,Zr)H$_4$ at 200 GPa shows a further suppressed $T_c$ of 69.55 K. %The exceptional superconducting properties of (Y,Sc)H$_4$ arise from the absence of a phonon gap between optical and acoustic branches and shorter bond lengths under pressure. 
The remarkable superconductivity in (Y,Sc)H$_4$ might be related to its unique phonon dispersion without optical-acoustic gap, compressed Y-H bonds, and significant electron delocalization under pressure, collectively boosting electron-phonon interactions.
Furthermore, the lowest optical phonons play a crucial role in the superconductivity of these materials. This work suggests that substituting Y with same-group metal elements is an effective strategy to enhance $T_c$ in hydride superconductors.

\end{abstract}
%---------------------------------------------------------------------

% PACS 2010 ALPHABETICAL INDEX
% https://publishing.aip.org/publishing/pacs/pacs-alphabetical-index
% PACS 2010 Regular Edition
% https://publishing.aip.org/publishing/pacs/pacs-2010-regular-edition
%\pacs{
%65.80.Ck,   %Thermal properties of graphene
%65.40.-b,    %Thermal properties of crystalline solids
%61.46.-w,    %Structure of nanoscale materials
%64.70.Nd,   %Structural transitions in nanoscale materials
%71.30.+h     %Metal-insulator transitions and other electronic transitions
%64.70.K-    %Solid-solid transitions
%78.67.Pt     %Multilayers; superlattices; photonic structures; metamaterials
%64.70.-p    %Specific phase transitions
%64.70.K-    %Solid-solid transitions
%73.22.-f,    %Electronic structure of nanoscale materials and related
             %systems
%73.20.At    %Surface states, band structure, electron density of
%            %states
%81.05.Cy    %Elemental semiconductors
%81.05.Zx     %New materials: theory, design, and fabrication
%81.16.-c    %Methods of micro- and nanofabrication and processing
%81.30.-t    %Phase diagrams and microstructures developed by
%            %solidification and solid-solid phase transformations
% }

% Insert suggested keywords - APS authors don't need to do this

% \maketitle must follow title, authors, abstract, \pacs, and \keywords
\maketitle

%---------------------------------------------------------------------
% \linenumbers\relax % Commence numbering lines

% If in two-column mode, this environment will change to single-column
% format so that long equations can be displayed. Use sparingly.
%\begin{widetext}
% put long equation here
%\end{widetext}

\section{I. Introduction}
%Superconducting materials have garnered a lot of interest and are widely acknowledged as being the core of the next industrial revolution. Among these, the H-series compounds have emerged as the hottest class of study objects in the field of superconductivity in recent years, given the high critical transition temperature ($T_c$) nearly approaches ambient temperature~\cite{Boeri_2022,10.1063/5.0077748,Du_2022}. The previous works in 2014-2015~\cite{drozdov2015conventional,duan2014pressure} found that H$_3$S has a remarkable $T_c$ of 203 K under roughly 90 GPa. For the Fm$\overline{3}$m structure of LaH$_{10}$, superconductivity at 170 GPa with a critical temperature of roughly 250 K was once again experimentally demonstrated in 2019~\cite{drozdov2019superconductivity}. Since then, other investigations have substantially expanded the family of H-based superconducting materials.

Superconducting materials have drawn significant attention and are pivotal to the next industrial revolution. Among these, hydrogen-rich compounds (H-series compounds) have emerged as a prominent focus in superconductivity research due to their exceptionally high critical transition temperatures ($T_c$), approaching ambient conditions~\cite{Boeri_2022,10.1063/5.0077748,Du_2022}. %Notably, studies from 2014-2015 revealed that H$_3$S exhibits a remarkable $T_c$ of 203 K under approximately 90 GPa
Notably, research conducted between 2014 and 2015 revealed that H$_3$S exhibited an exceptionally high critical temperature $T_c$ of 203 K at around 90 GPa, as confirmed by both experimental and theoretical studies~\cite{drozdov2015conventional,duan2014pressure}. %Further advancements came in 2019, with the Fm$\overline{3}$m structure of LaH$_{10}$ demonstrating superconductivity at 170 GPa and a $T_c$ of approximately 250 K confirmed through experiments~\cite{drozdov2019superconductivity}. %These breakthroughs have since stimulated extensive research, significantly expanding the family of hydrogen-based superconductors.
A major breakthrough came in 2019 when experiments verified that LaH$_{10}$ in the Fm$\overline{3}$m structure exhibits superconductivity at 170 GPa, reaching a $T_c$ of \~250 K.
These breakthroughs have spurred extensive research, significantly expanding the family of hydrogen-based superconductors through both experimental studies~\cite{bhattacharyya2024imaging,PhysRevLett.130.266001,SEMENOK202036,doi:10.1021/acs.inorgchem.1c01960,chen2024synthesis,kong2021superconductivity,https://doi.org/10.1002/adma.202006832,PhysRevLett.127.117001,nagamatsu2001superconductivity,minkov2022magnetic} and theoretical predictions~\cite{lucrezi2024full,Yao_2007,hou2015high,doi:10.1021/acs.jpcc.7b12124,doi:10.1021/acs.chemmater.5b04638,PhysRevLett.119.107001,doi:10.1073/pnas.2401840121,doi:10.1021/acs.jpclett.8b00615,doi:10.1021/jacs.2c05834,PhysRevB.102.144524}.

%A profusion of compounds with superconducting properties utilizing yttrium (Y) as a participating element has emerged in recent years among the numerous H-containing superconducting materials. For an instance, YH$_9$ in the $P$6/$mmm$ structure at 201 GPa has a $T_c$ of 243 K~\cite{kong2019superconductivity}. And YH$_6$ in the Im$\overline{3}$m structure at 166 GPa has a $T_c$ of 224 K~\cite{troyan2021anomalous}.
%at 201 GPa, the $T_c$ of YH$_9$ in the $P$6/$mmm$ structure was 243 K~\cite{kong2019superconductivity}. The $T_c$ of 224 K at 166 GPa was also discovered by Troyan et al. for the Im$\overline{3}$m structure of YH$_6$~\cite{troyan2021anomalous}. 
%Other materials that can achieve higher $T_c$ include the sodalite-like FCC YH$_{10}$, which can reach 305 $\sim$ 326 K at 250 GPa~\cite{liu2017potential}. Apart from the aforementioned binary hydrides, it is interesting to explore the exceptional superconducting characteristics of ternary hydrides involving Y. After stabilizing at 170 GPa, the cubic Fd$\overline{3}$m structure of CaYH$_{12}$ can display a high $T_c$ of 258 K at 200 GPa~\cite{liang2019potential}. Meanwhile, Pm$\overline{3}$m-YCaH$_{12}$ has an estimated $T_c$ at 180 GPa of 230 K and is stable at 200 GPa~\cite{xie2019high}.

In recent years, numerous hydrogen-containing superconducting materials featuring yttrium (Y) as a key element have been discovered, demonstrating remarkable superconducting properties. For instance, YH$_9$ with a $P$6/$mmm$ symmetry at 201 GPa exhibits a $T_c$ of 243 K~\cite{kong2021superconductivity}, while YH$_6$ in the Im$\overline{3}$m symmetry at 166 GPa achieves a $T_c$ of 224 K in the experimental research~\cite{https://doi.org/10.1002/adma.202006832}. Among materials with even higher $T_c$ values in theoretical research, the sodalite-like FCC YH$_{10}$ stands out, reaching 305–326 K at 250 GPa~\cite{liu2017potential}. Beyond these binary hydrides, ternary hydrides incorporating Y have also shown exceptional superconducting characteristics. For example, in theoretical research, the cubic Fd$\overline{3}$m structure of CaYH$_{12}$, stable at 170 GPa, demonstrates a $T_c$ of 258 K at 200 GPa~\cite{liang2019potential}. Similarly, Pm$\overline{3}$m-YCaH$_{12}$ is estimated to have a $T_c$ of 230 K at 180 GPa and remains stable at 200 GPa~\cite{xie2019high}.

%Nonetheless, it is evident that the ternary and binary hydrides previously discussed both need extraordinarily high pressure in order to attain superconductivity, and their applications are severely limited. %As a result, the approximate $T_c$ ranges of superconductors can only be obtained through simulations. 
%The key issue in the current application of H-based compound superconductors is how to reduce the pressure while maintaining a high $T_c$. Several studies have been conducted, for instance, it is discovered that although the $T_c$ of $\alpha$-MoB$_2$ at 90 GPa is 37 K, it can rise to approximately 43 K via electron doping~\cite{liu2022strong}. Additionally, the $T_c$ of two-dimensional hydrogenated MgB$_2$ can reach 100 K or more after applying biaxial tensile strains~\cite{PhysRevLett.123.077001}. And HCP-(La,Ce)H$_{9-10}$ has been efficiently synthesized and its high $T_c$ of 176 K at 100 GPa indicates its high-temperature superconductivity~\cite{chen2023enhancement}. While La and Ce are both rare-earth elements that are more costly to be doping at the same time, $\alpha$-MoB$_2$ itself does not have a high $T_c$, and studies for MgB$_2$ are not commonly employed as a bulk material.

Despite the promising superconducting properties of the aforementioned binary and ternary hydrides, their reliance on extraordinarily high pressures severely limits practical applications. %The key challenge in the development of hydrogen-based superconductors lies in reducing the required pressure while maintaining a high $T_c$. Several approaches have been explored to address this issue. 
The primary challenge in developing hydrogen-based superconductors is achieving high $T_c$ at reduced pressures, which has prompted exploration of multiple approaches.
For example, in theoretical research, the $T_c$ of $\alpha$-MoB$_2$, initially 37 K at 90 GPa, can be increased to approximately 43 K through electron doping~\cite{liu2022strong}. Similarly, two-dimensional hydrogenated MgB$_2$ achieves a $T_c$ exceeding 100 K under biaxial tensile strain in theoretical prediction~\cite{PhysRevLett.123.077001}. Additionally, HCP-(La,Ce)H$_{9-10}$, synthesized in the experiment at 100 GPa, exhibits a high $T_c$ of 176 K, demonstrating its potential for high-temperature superconductivity~\cite{chen2023enhancement}. 
%However, limitations remain: La and Ce, as rare-earth elements, are expensive for doping at the same time, $\alpha$-MoB$_2$ inherently possesses a low $T_c$, and the study of two-dimensional hydrogenation of MgB$_2$ is difficult to apply.
Nevertheless, several limitations remain: the prohibitive cost of using rare-earth elements (La, Ce) as dopants at the same time, the inherently low $T_c$ of $\alpha$-MoB$_2$, and the implementation difficulties associated with two-dimensional hydrogenation of MgB$_2$.

Building on the aforementioned discussion, we utilize YH$_4$ as the base material and create new compounds by substituting half of the Y atoms with Sc, La, and Zr, respectively. In this study, we determine the stabilization pressures for these three materials and employ calculations of differential charge density, Fermi surfaces, electronic band structures, and density of states to elucidate their electronic properties. Additionally, we investigate their phonon characteristics, including phonon dispersion and phonon density-of-state distribution. The $T_c$ ranges are determined using two distinct methods, with the effective Coulomb pseudopotential parameter ($\mu^*$) set to 0.05 and 0.4. Among the materials studied, (Y,Sc)H$_4$ at 100 GPa exhibits the highest $T_c$ and requires a lower synthesis pressure compared to the base material YH$_4$. Our findings demonstrate that substituting Y with same-group metal elements effectively enhances $T_c$, while substituting with neighboring elements significantly reduces the superconducting performance of material.

%In this work, on the basis of fully analysing the crystal structure, electrical and acoustic properties of (Y,Sc)H$_4$(100 GPa), (Y,La)H$_4$(120 GPa) and (Y,Zr)H$_4$(200 GPa), found that when the effective Coulomb pseudopotential parameters $\mu^*$ were taken to be 0.1, the $T_c$ of (Y,Sc，La,Zr)H$_4$ was calculated by using three different procedures, and the $T_c$ of (Y,Sc)H$_4$(100 GPa) was calculated to be between 104.363 $\sim$ 143 K and 82.486 $\sim$ 128 K for (Y,La)H$_4$(120 GPa). By substituting 50 per cent of Y, the $T_c$ of (Y,Zr)H$_4$(200 GPa) falls between 52.611 $\sim$ 100 K. 

\section{II. COMPUTATIONAL METHODS}
%The exact optimization of the lattice constants of the calculated materials was achieved by using the Vienna Ab-initio Simulation Package (VASP)~\cite{KRESSE199615,PhysRevB.54.11169} with the Perdew-Burke-Ernzerhof (PBE)~\cite{PhysRevLett.77.3865} Generalized Gradient Approximation (GGA)~\cite{PhysRevB.50.4954,PhysRevB.73.235116}. This allowed for the calculation and analysis of their electronic properties, which included density of states, electronic bands, and differential charge density, as well as their phonon properties, which included phonon spectrum and phonon density of states. %In the projector augmented wave (PAW)~\cite{KRESSE199615,PhysRevB.48.13115,PhysRevB.50.17953} framework, the bare ion Coulomb potential was addressed. 
%For the electron self-consistent and non-self-consistent computations, the \textit{k}-point grid was chosen to be 12 $\times$ 12 $\times$ 6 and the energy cutoff was set to 350 eV. The structures were fully optimized when the maximum energy and force were less than 10$^{-6}$ eV/\AA ~ and 1 meV/\AA ~, respectively.

The lattice constants of the calculated materials were optimized using the Vienna $ab$-$initio$ Simulation Package (VASP)~\cite{KRESSE199615,PhysRevB.54.11169}, employing the Perdew-Burke-Ernzerhof (PBE)~\cite{PhysRevLett.77.3865} Generalized Gradient Approximation (GGA)~\cite{PhysRevB.50.4954,PhysRevB.73.235116}. This optimization facilitated the calculation and analysis of electronic properties, including density of states, electronic band structures, and differential charge density, as well as phonon properties, such as the phonon spectrum and phonon density of states. For both self-consistent and non-self-consistent electron computations, a \textit{k}-point grid of 12 $\times$ 12 $\times$ 6 was used, with an energy cutoff of 350 eV. Structural optimization was considered complete when the maximum energy and force converged to below 10$^{-6}$ eV/\AA~and 1 meV/\AA~, respectively.

%Utilizing the optimized norm-conserving Van derbilt (ONCV)~\cite{van2018pseudodojo} 
%pseudopotentials with the density functional perturbation theory (DFPT) method, the Quantum-ESPRESSO (QE)~\cite{Giannozzi_2009} package with the charge density cutoff of 1040 Ry and kinetic energy cutoff of 130 Ry, as well as \textit{k}-point and \textit{q}-point grids of 12 $\times$ 12 $\times$ 6 and 6 $\times$ 6 $\times$ 3, respectively, allows for the computation of the electron-phonon coupling (EPC), the spectral function and superconducting properties.
%Additionally, the EPW~\cite{Giannozzi_2017} program for the anisotropic Migdal-Eliashberg equations~\cite{eliashberg1960interactions,PhysRevB.87.024505} and the ELK code~\cite{elk} based on the isotropic Migdal-Eliashberg equations~\cite{FLORESLIVAS20201,10.1063/5.0077748} were employed.

The electron-phonon coupling (EPC), spectral function, and superconducting properties were computed using the Quantum-ESPRESSO (QE)~\cite{Giannozzi_2009} package with optimized norm-conserving Vanderbilt (ONCV)~\cite{van2018pseudodojo} pseudopotentials and the density functional perturbation theory (DFPT) method. Calculations utilized a charge density cutoff of 1040 Ry, a kinetic energy cutoff of 130 Ry, and \textit{k}-point and \textit{q}-point grids of 12 $\times$ 12 $\times$ 6 and 6 $\times$ 6 $\times$ 3, respectively. %Additionally, the EPW~\cite{Giannozzi_2017} program was used to solve the anisotropic Migdal-Eliashberg equations~\cite{PhysRevB.87.024505}, while the ELK code~\cite{elk} was employed for calculations based on the isotropic Migdal-Eliashberg equations~\cite{FLORESLIVAS20201,10.1063/5.0077748}.
The isotropic Migdal–Eliashberg equations were solved using the EPW program~\cite{Giannozzi_2017,PhysRevB.87.024505} and the ELK code~\cite{elk,FLORESLIVAS20201,10.1063/5.0077748}.

With the effective Coulomb pseudopotential parameters $\mu^*$ ranging from 0.05 to 0.40, the McMillan-Allen-Dynes formula~\cite{PhysRevB.12.905} is utilized to calculate the $T_c$ of (Y,Sc,Zr,La)H$_4$,

%The McMillan-Allen-Dynes formula~\cite{PhysRevB.12.905} was employed to calculate the $T_c$ of (Y,Sc,Zr,La)H$_4$, with effective Coulomb pseudopotential parameters ($\mu^*$) ranging from 0.05 to 0.13.

%
\begin{eqnarray}
\label{eqn1}	 
{T}_{c} = \frac{ \omega_{\rm log}}{1.2} \exp \left[-\frac{1.04(1+\lambda)}{\lambda-\mu^*(1+0.62\lambda)}\right],
\end{eqnarray}
%

%The effective Coulomb pseudopotential, denoted as $\mu^*$, has a standard value of 0.10. 
The electron-phonon coupling constant integrated over an integrated domain is,

\begin{eqnarray}
\label{eqn2}	 
\lambda(\omega)= 2 \int_0^\omega \frac{\alpha^2 F(\omega')}{\omega'} d\omega',
\end{eqnarray}
where $\lambda(\omega\rm_{max})$ is the maximum phonon frequency and $\lambda$ is the electron-phonon coupling constant $\lambda$ utilized in Eq.~(\ref{eqn1}). Moreover, the characteristic phonon frequency $\omega\rm_{log}$, which is logarithmically averaged, can be expressed as,

\begin{eqnarray}
\label{eqn3}	 
\omega_{\rm log} = \exp \left[\frac{2}{\lambda} \int \frac{d\omega}{\omega} \alpha^2 F(\omega) \ln\omega \right],
\end{eqnarray}

The following are the isotropic Migdal-Eliashberg equations~\cite{FLORESLIVAS20201,10.1063/5.0077748},

\begin{eqnarray}
\begin{split}
\label{eqn4}	 
Z(i\omega_n) =  1 + \frac{\pi T}{\omega_n} \sum_{n'} \frac{\omega_{n'}}{\sqrt{\omega_{n}^2 +\Delta^2 (i\omega_{n})}} \times \lambda(n-n'),
\end{split}
\end{eqnarray}
\begin{eqnarray}
\begin{split}
\label{eqn5}	 
Z(i\omega_n)\Delta(i\omega_n) =  &{\pi T} \sum_{n'} \int d\xi\,
                                 \frac{\Delta(i\omega_{n'})}{\sqrt{\omega_{n'}^2 +\Delta^2 (i\omega_{n'})}} \\
                                 &\times\left[\lambda(n-n')-\mu^*\right],
\end{split}
\end{eqnarray}

where the superconducting gap $\Delta(i\omega_n)$ and the fermionic Matsubara frequencies $\omega_{n'}$ are represented, together with the renormalization function $Z(i\omega_n)$. Consequently, the definition of the Eliashberg spectral function is,

\begin{eqnarray}
\label{eqn6}	 
\alpha^2 F(\omega) = \frac{1}{2 \pi N(0)} \sum_{qj} \frac{\gamma_{qj}}{\omega_{qj}} \delta(\hbar \omega - \hbar \omega_{qj}),
\end{eqnarray}
where $\omega_{qj}$ denotes frequency with $a$ phonon $j$ and a wave vector $q$, and $\gamma_{qj}$ denotes phonon linewidth.

\begin{figure}
\includegraphics[width=1.0\columnwidth]{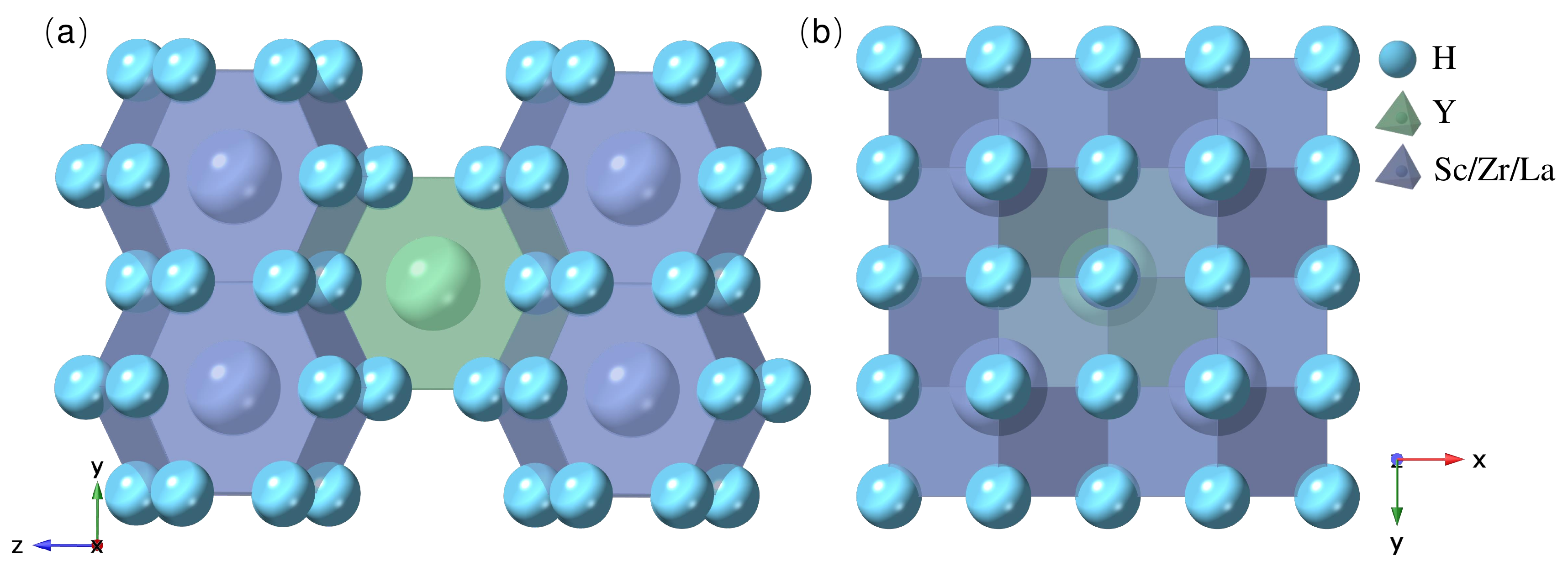}
%\vspace{-2mm}
\caption{%
%The crystal structure representations of (Y,Sc)H$_4$ at 100 GPa, (Y,Zr)H$_4$ at 200 GPa and (Y,La)H$_4$ at 120 GPa. The structure of (Y,Sc,Zr,La)H$_4$ is shown in (a) front and (b) side views, where the blue, green, purple spheres stand for element H, element Y, element Sc, Zr or La, respectively.
%
Crystal structure representations of (Y,Sc)H$_4$ at 100 GPa, (Y,Zr)H$_4$ at 200 GPa, and (Y,La)H$_4$ at 120 GPa. The structures of (Y,Sc,Zr,La)H$_4$ are depicted in (a) front and (b) side views. Blue spheres represent hydrogen (H), while green, purple, and other colored spheres represent the metal elements Y, Sc, Zr, or La, respectively.
\label{fig1}}
\end{figure}
%\end{figure*}
%===========< FIGURE 1 >=========================================

%===========< FIGURE 2 >=========================================
%\begin{figure*}
\begin{figure}
\includegraphics[width=1.0\columnwidth]{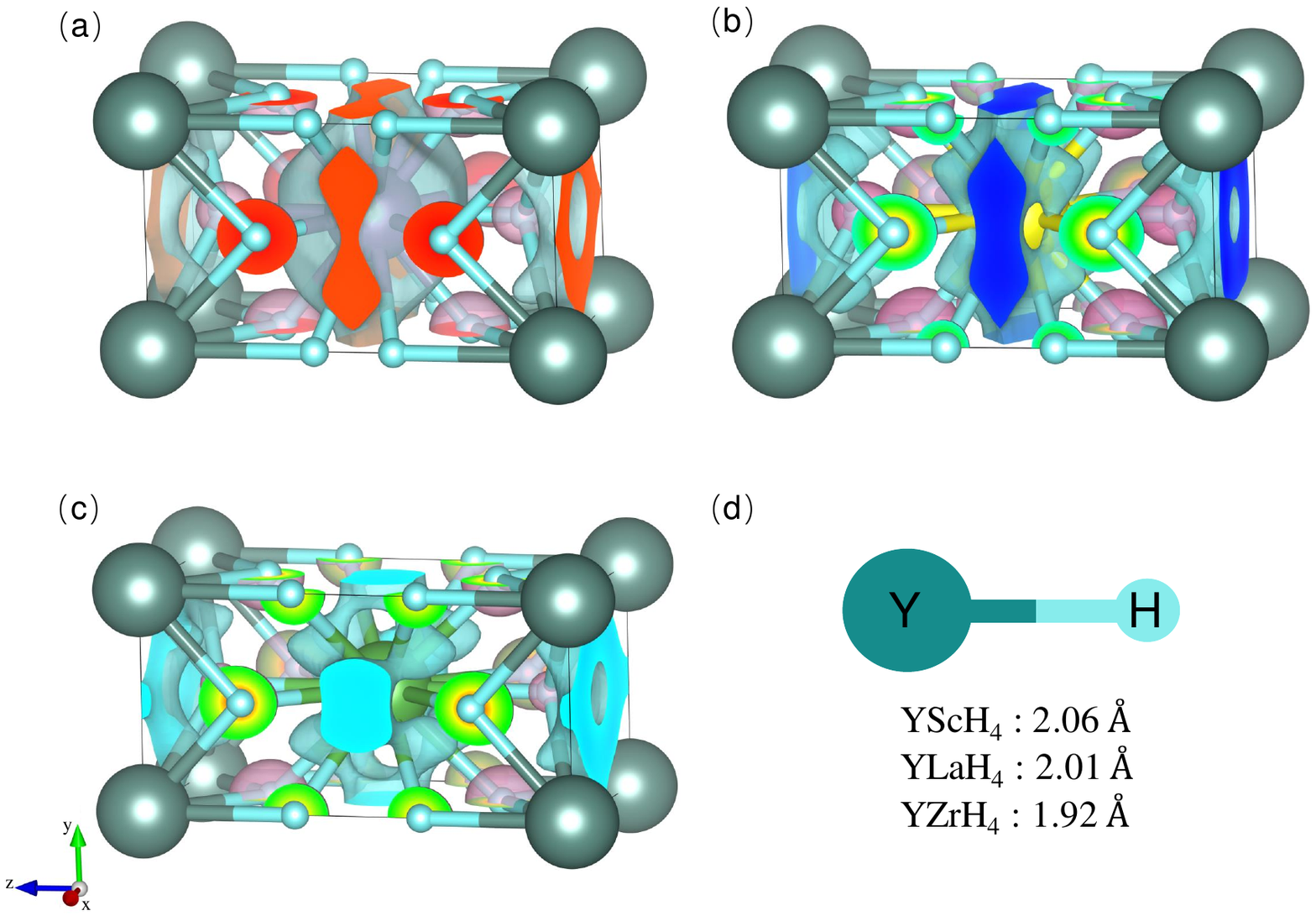}
%\vspace{-2mm}
\caption{%
%The differential charge density diagram for (Y,Sc)H$_4$, (Y,Zr)H$_4$ and (Y,La)H$_4$. The front view of differential charge densities of (a) (Y,Sc)H$_4$, (b) (Y,Zr)H$_4$ and (c) (Y,La)H$_4$, where the small light blue sphere represents the element H, the large dark cyan sphere represents the element Y, and the large pink, yellow and light green spheres in the center represent Sc, Zr and La, respectively. Additionally, the negative regions are shown in blue and the positive regions are shown in pink. (d) The average bond lengths of elements Y and H in three materials.
%
The front views of the differential charge densities are shown for (a) (Y,Sc)H$_4$, (b) (Y,Zr)H$_4$, and (c) (Y,La)H$_4$. Small light blue spheres represent hydrogen (H), large dark cyan spheres represent yttrium (Y), and the large pink, yellow, and light green spheres at the center represent Sc, Zr, and La, respectively. Negative charge regions are depicted in blue, while positive charge regions are shown in pink. (d) The average bond lengths between Y and H elements for the three materials.
\label{fig2}}
%\end{figure*}
\end{figure}
%===========< FIGURE 2 >=========================================

%===========< FIGURE 3 >=========================================
\begin{figure*}
%\begin{figure}
\includegraphics[width=2.0\columnwidth]{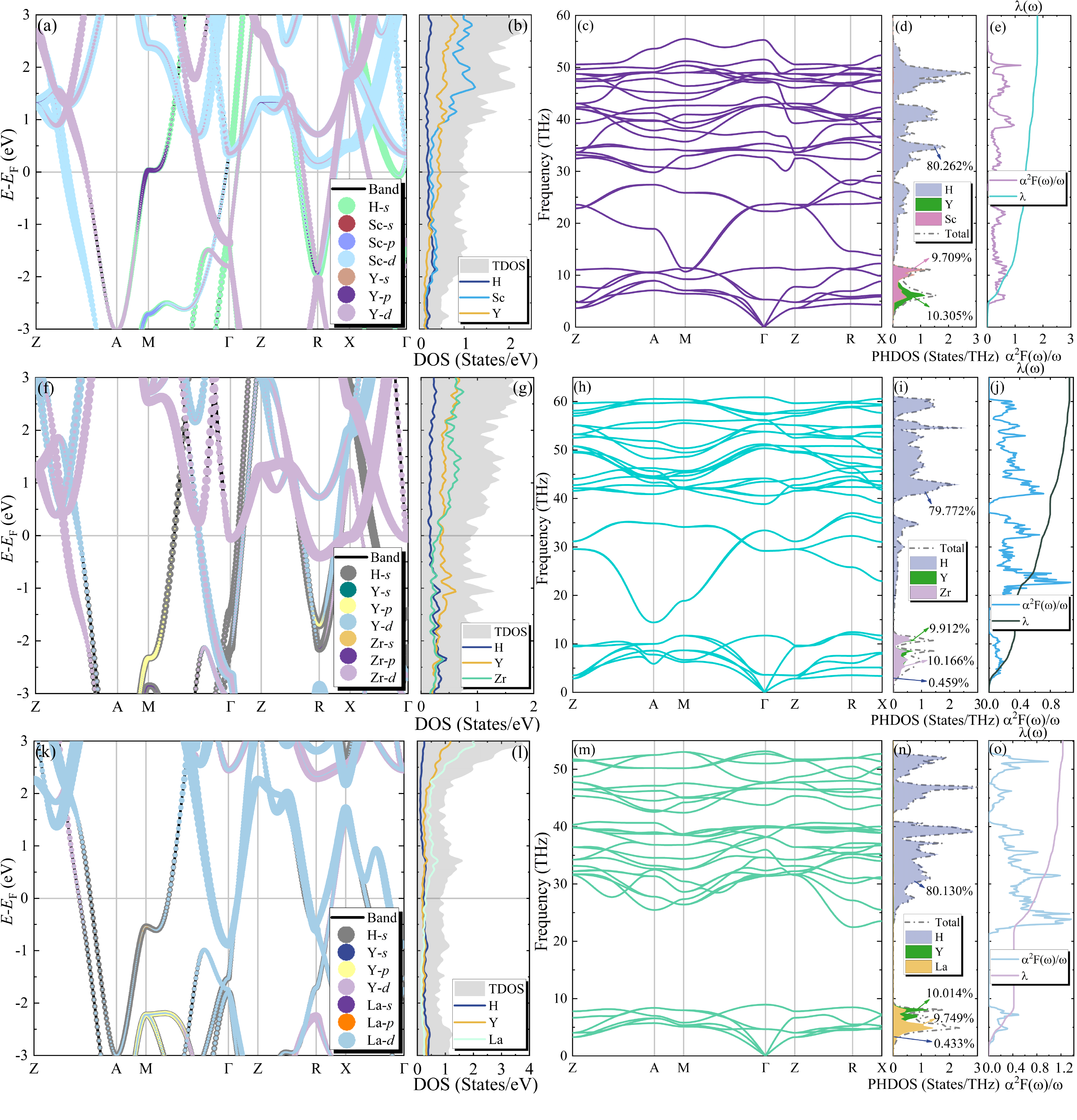}
%\includegraphics[width=\columnwidth]{fig4}
%\vspace{-2mm}
\caption{
The electronic and phonon properties of (Y,Sc)H$_4$ (100 GPa), (Y,Zr)H$_4$ (200 GPa), and (Y,La)H$_4$ (120 GPa), respectively.
(a) The electronic band structures and (b) the electronic density of states (EDOS) for (Y,Sc)H$_4$. The gray-scale highlights orbital contributions, with H, Sc and Y orbitals represented by dark blue, light blue and orange lines, respectively. (c) The phonon dispersion and (d) phonon density of states (PDOS) for (Y,Sc)H$_4$, where H, Y, and Sc are represented by dark blue, green, and pink shades. The total PDOS is shown as a gray dotted line, with arrows indicating the percentage of PDOS contributed by each element. (e) The electron-phonon coupling integral $\lambda$($\omega$) (aqua line) and the Eliashberg spectral function $\alpha^2$F ($\omega$) (violet line). (f–j) and (k–o) depict the corresponding properties for (Y,Zr)H$_4$ and (Y,La)H$_4$, respectively.
\label{fig3}}
%\end{figure}
\end{figure*}
%===========< FIGURE 3 >=========================================

\section{III. RESULTS AND DISCUSSION}
As the matrix material investigated in this study, YH$_4$ crystallizes in the $I$4/$mmm$ space group (No. 139), exhibiting an experimental superconducting transition temperature ($T_c$) of \textasciitilde88 K at 155 GPa~\cite{PhysRevB.104.174509}, while theoretical predictions estimate $T_c$ between 84-95 K at 120 GPa~\cite{li2015pressure}. The (Y,Sc)H$_4$, (Y,Zr)H$_4$ and (Y,La)H$_4$ compound maintains structural stability at 100, 200, and 120 GPa respectively, crystallizing in the same space group as YH$_4$, as shown in Fig.~\ref{fig1}(a-b). Complete structural modeling parameters and thorough analyses of both kinetic and thermodynamic stability for (Y,Sc)H$_4$ (100 GPa), (Y,Zr)H$_4$ (200 GPa), and (Y,La)H$_4$ (120 GPa) systems are presented in Fig.S3 and S2. The optimized lattice constants are as follows: for (Y,Sc)H$_4$, a = b = 2.856 \AA, c = 5.114 \AA; for (Y,Zr)H$_4$, a = b = 2.678 \AA, c = 5.056 \AA; for (Y,La)H$_4$, a = b = 2.811 \AA, c = 5.648 \AA. Each structure consists of a cage of 18 hydrogen atoms enclosing the heavier metal atoms (Y, Sc, La, or Zr). These materials share the same structure as other MH$_4$ compounds (M representing different metal elements)~\cite{https://doi.org/10.1002/chem.202102679}, including pure YH$_4$~\cite{li2015pressure,PhysRevB.104.174509}, ScH$_4$~\cite{doi:10.1021/acs.jpcc.7b12124,PhysRevB.96.144108,PhysRevB.96.094513,https://doi.org/10.1002/chem.202102679}, ZrH$_4$~\cite{https://doi.org/10.1002/chem.202102679,PhysRevB.98.134103}, and LaH$_4$~\cite{liu2017potential}. Interestingly, hydrogen compounds composed of Y (group IIIB) and Zr (group IVB) require higher pressures to stabilize, despite their similar elemental weights. However, hydrides with elements Y, Sc, or La from group IIIB can stabilize at comparatively lower pressures.

Figure~\ref{fig2} shows the differential charge density and average Y–H bond lengths for (Y,Sc)H$_4$, (Y,La)H$_4$ and (Y,Zr)H$_4$, while Fig. S4 presents the electron localization function (ELF) maps along the (110) plane. Interestingly, we find that some electrons detached from the central atoms (Sc, Zr, and La) become highly localized in the interstitial regions near hydrogen sites. This electron distribution closely resembles that of the near-room-temperature superconductor LaH$_{10}$~\cite{PhysRevMaterials.5.024801}, suggesting a possible link to high-$T_c$  behavior. Furthermore, distinct bonding characteristics are observed across the three compounds, which critically influence their electronic structures and likely govern their superconducting properties.

(Y,Sc)H$_4$, stabilized at a relatively low pressure of 100 GPa, exhibits the strongest charge redistribution and highly localized covalent Y–H bonds, with ELF values approaching 1 and the longest average Y–H bond length (2.06 \AA). This indicates that the strong electron-phonon coupling is closely associated with the material’s bonding framework, which is enhanced under compression, rather than being purely a consequence of external pressure.
%This indicates that its strong electron-phonon coupling originates primarily from intrinsic bonding characteristics, rather than external compression, an advantageous feature for achieving high-temperature superconductivity under moderate conditions.
In contrast, (Y,Zr)H$_4$ requires a much higher pressure of 200 GPa to stabilize, yet shows the weakest charge redistribution and lowest ELF values. Its shortest Y–H bond length (1.92 \AA) suggests a more compact but less covalent bonding environment. The structural rigidity introduced by Zr likely suppresses bond flexibility and electron localization, thus weakening electron-phonon coupling.

At an intermediate stabilization pressure of 120 GPa, (Y,La)H$_4$ demonstrates intermediate charge redistribution characteristics accompanied by partial electron localization, featuring an average Y-H bond distance of 2.01 \AA. The larger ionic radius and lower electro-negativity of La may lead to a more relaxed bonding environment that offsets pressure-induced bond compression.
The superconducting potential of these hydrides depends more critically on the nature of hydrogen-mediated covalent bonding and electron localization than on the applied pressure. Among them, (Y,Sc)H$_4$ stands out as a promising candidate due to its favorable bonding characteristics maintained even at low pressure.

Figure~\ref{fig3}(a), (f), and (k) depict the electronic band structures with orbital resolution of (Y,Sc)H$_4$(100 GPa), (Y,Zr)H$_4$(200 GPa) and (Y,La)H$_4$(120 GPa), respectively. Combined with the electronic density of states (EDOS) for (Y,Sc)H$_4$ at 100 GPa, as shown in Fig.~\ref{fig3}(b), the total EDOS value is approximately 1 state/eV. The contributions from various orbitals H-$s$, Sc-$s$, Sc-$p$, Sc-$d$, Y-$s$, Y-$p$, and Y-$d$ are represented by light-green, dark-red, dark-blue, light-bluish, dark-yellowish, dark-violet, and mauve dots, respectively. Moreover, Fig.~\ref{fig3}(a) reveals that the electronic bands near the Fermi level for (Y,Sc)H$_4$ are predominantly composed of Y-$d$, Sc-$d$, and the H-$s$ orbitals. 

%Because La and Sc are in the group $\textup{III}$B, the electronic bands for Fig.~\ref{fig4}(k) and Fig.~\ref{fig4}(a) are similar. Sc has the 3d$^1$ valence electrons, while La has the 5d$^1$ valence electrons. Therefore, (Y,La)H$_4$ has a higher Fermi level compare to the (Y,Sc)H$_4$. However, the electronic bands of (Y,Zr)H$_4$, shown in Fig.~\ref{fig4}(f), are significantly different from those of the two materials mentioned above because Zr has the valence electron of 4d$^2$ that is one electron more than Sc and La, which are members of the same $\textup{III}$B group. 

Since both La and Sc belong to group $\textup{III}$B, the electronic bands in Fig.~\ref{fig3}(k) and Fig.~\ref{fig3}(a) are similar. Sc has 3d$^1$ valence electrons, while La has 5d$^1$ valence electrons, resulting in a higher Fermi level for (Y,La)H$_4$ compared to (Y,Sc)H$_4$. In contrast, the electronic bands of (Y,Zr)H$_4$, depicted in Fig.~\ref{fig3}(f), differ significantly from the other two materials. This is because Zr, a member of group $\textup{IV}$B, has 4d$^2$ valence electrons meaning one more than the valence electrons of Sc and La.

%And Fig.~\ref{fig4}(f) shows the orbits of H-$s$, Y-$s$, Y-$p$, Y-$d$, Zr-$s$, Zr-$p$, and Zr-$d$, which are represented by dark grey, dark green, yellow, light blue, orange, dark purple, and mauve lines, respectively. 
%Fig.~\ref{fig4}(f) shows the orbits of different elements around Fermi level.
%It is found that Zr-$d$, Y-$d$, and H-$s$ orbitals make up the electronic bands of (Y,Zr)H$_4$ in 200 GPa at the Fermi level of (Y,Zr)H$_4$. Similarly, the electronic bands near the E$_F$ of (Y,La)H$_4$ consists mainly of H-$s$, La-$d$ and a small number of Y-$d$ orbitals shown in Fig.~\ref{fig4}(k). As can be shown in Fig.~\ref{fig4}(g) and Fig.~\ref{fig4}(l), the E$_F$ of (Y,Zr)H$_4$ and (Y,La)H$_4$ has a total EDOS value of around 0.9 states/eV and 0.94 states/eV, respectively. However, they are less than that of (Y,Sc)H$_4$. Therefore, the high $T_c$ of (Y,Sc)H$_4$ can be distinguished by the amplitude of the EDOS value around the E$_F$.

%As seen in Fig.~\ref{fig4}(k), the electronic bands of (Y,La)H$_4$ is composed of dark grey H-$s$, dark blue Y-$s$, yellow Y-$p$, mauve Y-$d$, dark purple La-$s$ orbitals, orange La-$p$, and light blue La-$d$ orbitals. 
%with the van Hove singularity around 1 eV under the E$_F$ shown in Fig.~\ref{fig4}(k)
% The EDOS at the E$_F$ is 0.94 states/eV shown in Fig.~\ref{fig4}(l), which is also lower than the (Y,Sc)H$_4$. 

Figure~\ref{fig3}(f) illustrates the orbital contributions near the Fermi level for (Y,Zr)H$_4$ at 200 GPa, where Zr-$d$, Y-$d$, and H-$s$ orbitals dominate the electronic bands. Similarly, the electronic bands near the Fermi level for (Y,La)H$_4$, as shown in Fig.~\ref{fig3}(k), are primarily composed of H-$s$, La-$d$, and a smaller contribution from Y-$d$ orbitals. Fig.~\ref{fig3}(g) and (l) show that the total electronic density of states (EDOS) at Fermi level (E$_F$) is approximately 0.9 states/eV for (Y,Zr)H$_4$ and 0.94 states/eV for (Y,La)H$_4$. Both values are lower than the EDOS of (Y,Sc)H$_4$, which accounts for its higher $T_c$. The enhanced $T_c$ of (Y,Sc)H$_4$ is thus correlated with the greater EDOS amplitude around E$_F$. 

%The phonon dispersions of (Y,Sc)H$_4$, (Y,Zr)H$_4$ and (Y,La)H$_4$ are displayed in Fig.~\ref{fig4}(c), Fig.~\ref{fig4}(h) and Fig.~\ref{fig4}(m), which demonstrate the dynamical stability of these materials. It can be noticed that the gap between the optical and acoustic branches in Fig.~\ref{fig4}(c), Fig.~\ref{fig4}(h) and Fig.~\ref{fig4}(m) is getting bigger. Among these phonon branches, the element H easily dominates the optical branch at high frequencies, while the metallic elements Y, Sc, La and Zr in three materials are all distributed in the acoustic branch at low frequencies, as shown in Fig.~\ref{fig4}(d), Fig.~\ref{fig4}(i) and Fig.~\ref{fig4}(n). Sc, Zr and La occupy almost the same proportion of the phonon density of states (PDOS). Further calculations were made for (Y,Sc)H$_4$,(Y,Zr)H$_4$ and (Y,La)H$_4$, resulting in the Eliashberg spectral function $\alpha^2$F($\omega$) and electron-phonon coupling $\lambda$($\omega$), which are displayed in Fig.~\ref{fig4}(e), Fig.~\ref{fig4}(j) and Fig.~\ref{fig4}(o).

The phonon dispersions of (Y,Sc)H$_4$, (Y,Zr)H$_4$, and (Y,La)H$_4$ are presented in Fig.~\ref{fig3}(c), (h), and (m), respectively, demonstrating the dynamical stability of these materials. Notably, the gap between the optical and acoustic branches increases across Fig.~\ref{fig3}(c), (h) and (m). At high frequencies, hydrogen dominates the optical branches, while the metallic elements Y, Sc, La, and Zr contribute predominantly to the acoustic branches at low frequencies, as illustrated in Fig.~\ref{fig3}(d), (i) and (n). The phonon density of states (PDOS) shows that Sc, Zr, and La occupy nearly the same proportion. Further calculations for (Y,Sc)H$_4$, (Y,Zr)H$_4$, and (Y,La)H$_4$ yield the Eliashberg spectral function $\alpha^2$F($\omega$) and the electron-phonon coupling $\lambda$($\omega$), depicted in Fig.~\ref{fig3}(e), (j) and (o).

\begin{figure*}
%\begin{figure}
\includegraphics[width=2.0\columnwidth]{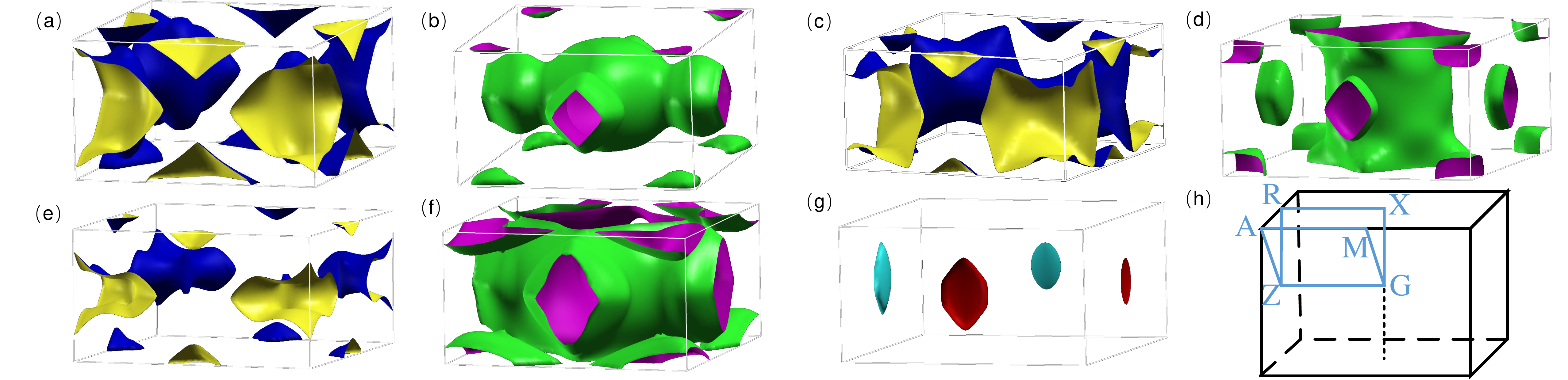}
%\includegraphics[width=1.0\columnwidth]{fig3}
%\vspace{-2mm}
\caption{
%The calculated Fermi surface from (a) the 1st energy band, (b) the 2nd energy band in (Y,Sc)H$_4$. (c) and (d) represent the 1st and 2nd energy bands in (Y,La)H$_4$, respectively. In addition, from (e) the 1st energy band, (f) the 2nd energy band as well as (g) the 3rd energy band in (Y,Zr)H$_4$ generated by xcrysden~\cite{KOKALJ1999176}, with the $k$-path consisting of (h) high symmetry points Z, A, M, $\Gamma$, R and X by blue lines. 
%
Calculated Fermi surfaces for (Y,Sc)H$_4$, (Y,La)H$_4$, and (Y,Zr)H$_4$. Panels (a) and (b) show the 1st and 2nd energy bands of (Y,Sc)H$_4$, while panels (c) and (d) depict the 1st and 2nd energy bands of (Y,La)H$_4$. Panels (e), (f), and (g) illustrate the 1st, 2nd, and 3rd energy bands of (Y,Zr)H$_4$, respectively, as generated using Xcrysden~\cite{KOKALJ1999176}. (h) The $k$-path, comprising high-symmetry points Z, A, M, $\Gamma$, R, and X, is represented by blue lines in panel.
\label{fig4}}
%\end{figure}
\end{figure*}

Interestingly, the electron-phonon coupling strength $\lambda$ and spectral functions for (Y,Zr)H$_4$ and (Y,La)H$_4$ indicate that high-frequency phonons play a critical role in the superconducting properties of both materials. These high-frequency phonons, represented by the optical branches, are predominantly contributed by the hydrogen (H) atoms, as shown in Fig.~\ref{fig3}(h) and (m). In contrast, an important distinction for (Y,Sc)H$_4$ is the absence of a gap between the acoustic and optical branches, as observed by comparing Fig.~\ref{fig4}(c) and (e). The $\lambda$ values for (Y,Sc)H$_4$, (Y,Zr)H$_4$, and (Y,La)H$_4$ are 1.80, 1.05, and 1.23, respectively, further highlighting the enhanced electron-phonon coupling in (Y,Sc)H$_4$.

%===========< FIGURE 4 >=========================================
%===========< FIGURE 4 >=========================================

%===========< FIGURE 5 >=========================================
%\begin{figure*}
\begin{figure}
\includegraphics[width=1.0\columnwidth]{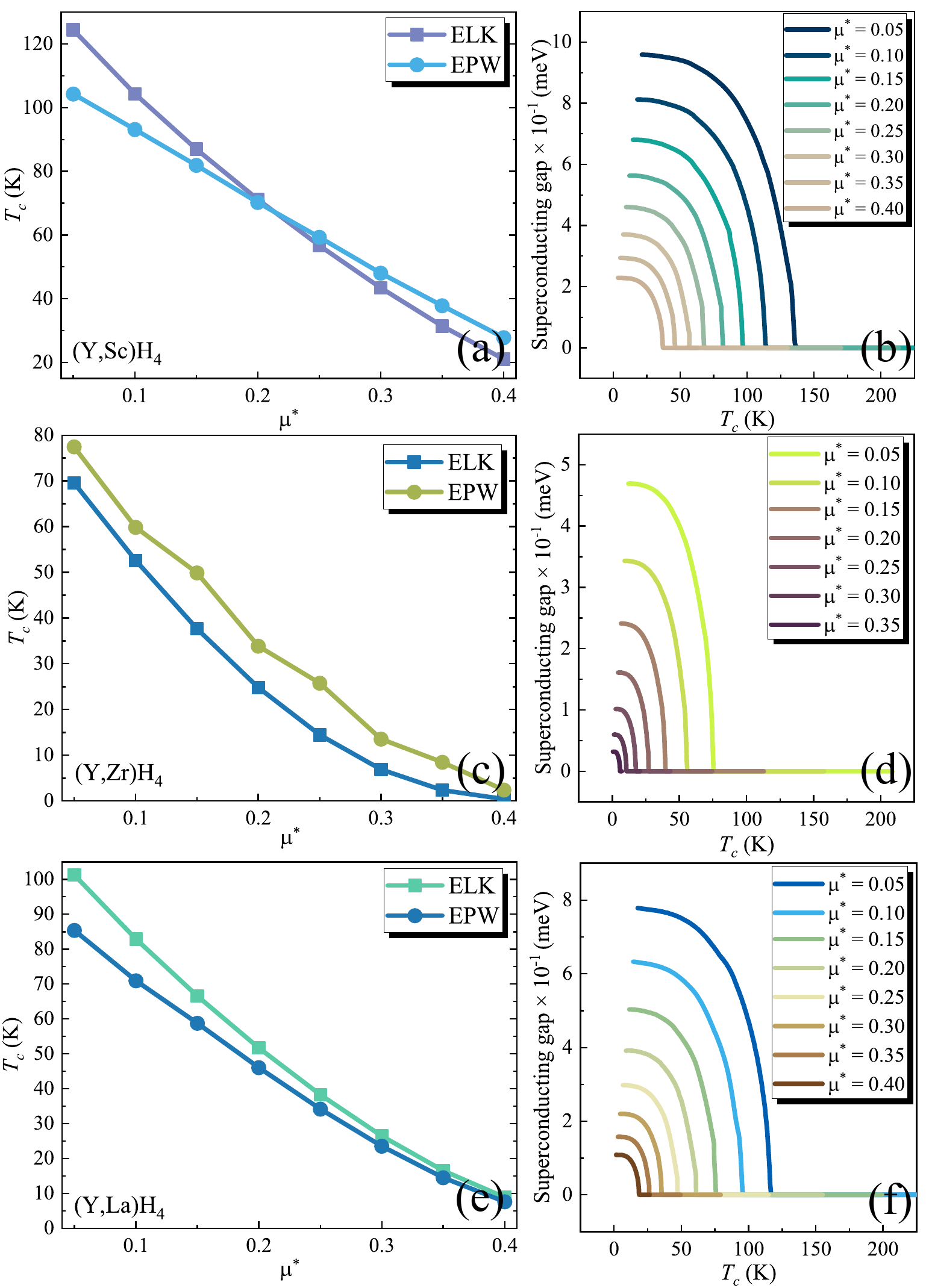}
%\includegraphics[width=\columnwidth]{fig5}
%\vspace{-2mm}
\caption{
%Superconducting characteristics of (Y,Sc)H$_4$, (Y,Zr)H$_4$ and (Y,La)H$_4$. The dark purple, yellow, and green curves in panels (a), (d), and (g) represent the calculated $T_c$ values using the QE code~\cite{Giannozzi_2009} with $\mu^*$ ranging from 0.05 to 0.40, while the blue, navy blue, and denim blue curves in the same panels correspond to the Tc values computed using the ELK code~\cite{elk}, as detailed in panels (b), (e), and (h), respectively. (c), (f), and (i) depict the $T_c$ values for (Y,Sc)H$_4$, (Y,Zr)H$_4$ and (Y,La)H$_4$, respectively, calculated using the EPW code~\cite{Giannozzi_2017} with $\mu^*$=0.10.
Superconducting characteristics of (Y,Sc)H$_4$, (Y,Zr)H$_4$ and (Y,La)H$_4$. The dark purple, yellow, and green curves in panels (a), (d), and (g) represent the calculated $T_c$ values using the ELK code~\cite{elk} with $\mu^*$ ranging from 0.05 to 0.40, while the blue, navy blue, and denim blue curves in the same panels correspond to the $T_c$ values computed using the EPW code~\cite{Giannozzi_2017}, as detailed in panels (b), (d), and (f), respectively.
\label{fig5}}
\end{figure}

As shown in Fig.~\ref{fig4}, the Fermi surface (FS) topologies of (Y,Sc)H$_4$, (Y,Zr)H$_4$, and (Y,La)H$_4$ differ markedly, reflecting distinct orbital contributions and symmetry-derived band degeneracies. These topological features directly correlate with the band structures and electron–phonon coupling constants ($\lambda$) extracted from Fig.~\ref{fig3}, highlighting the critical role of electronic structure in determining superconducting properties.
In (Y,Sc)H$_4$, the Fermi surface consists of a three-dimensional, highly interconnected network formed by degenerate $E_g$ bands, primarily originating from Y or Sc-$d$ orbitals crossing the Fermi level along high-symmetry directions such as $\Gamma$–Z and R–A. This topology promotes strong electronic states overlap and scattering phase space, resulting in a substantial electron–phonon coupling constant of $\lambda$ = 1.80, indicative of strong superconducting potential.

In contrast, (Y,La)H$_4$ exhibits a quasi-two-dimensional Fermi surface composed of cylindrical sheets primarily derived from the $d$ orbitals of Y, with minimal dispersion along the $k_z$ direction. Although the band remains degenerate, the reduced interlayer electronic overlap and stronger anisotropy yield a moderately strong coupling of $\lambda$ = 1.23.
Meanwhile, (Y,Zr)H$_4$ presents a collection of discrete, nearly spherical and spindle-shaped Fermi pockets spread across the Brillouin zone, arising from hybridized Zr-$d$ and Y or Zr-$d$ orbitals. This fragmented and weakly connected Fermi surface leads to a lower density of states at the Fermi level and a relatively weak electron–phonon interaction, with $\lambda$ = 1.05.
The dimensionality, degeneracy, and multiplicity of Fermi surface sheets substantially modulate the electronic density of states and electron–phonon coupling strength, ultimately governing the superconducting transition temperature achievable in each compound. Additionally, Fig.~\ref{fig3}(h) highlights the high-symmetry point pathways for the three materials.

%\subsection{C. Superconductivity properties}
%Three methods are used to systematically determine the superconducting properties of these three materials, displayed in Fig.~\ref{fig5}. The computational details are shown in the Eq.\eqref{eqn1}, Eq.\eqref{eqn4} and \eqref{eqn5}, Eq.\eqref{eqn7} and \eqref{eqn8}, respectively. Utilizing the QE code~\cite{Giannozzi_2009}, the $T_c$ with $\mu^*$ was determined in Fig.~\ref{fig5}(a), Fig.~\ref{fig5}(d) and Fig.~\ref{fig5}(g). It is intuitively found that the $T_c$ decreases linearly and monotonically as the $\mu^*$ increases in the range of 0.05 to 0.13. In the following, we choose $\mu^*$=0.10 to study the $T_c$ of (Y,Sc)H$_4$, (Y,Zr)H$_4$ and (Y,La)H$_4$. 

Two methods were employed to systematically determine the superconducting properties of (Y,Sc)H$_4$, (Y,Zr)H$_4$, and (Y,La)H$_4$, as shown in Fig.~\ref{fig5}. The computational details are outlined in Eqs. \eqref{eqn1}, \eqref{eqn4} and \eqref{eqn5}. Using the EPW code~\cite{Giannozzi_2017}, the critical temperature $T_c$ as a function of the Coulomb pseudopotential parameter $\mu^*$ was calculated, with results displayed in Fig.~\ref{fig5}(a), (c) and (e). To investigate the $T_c$ of (Y,Sc)H$_4$, (Y,Zr)H$_4$, and (Y,La)H$_4$ systems, we selected $\mu^*$ = 0.05 in our EPW calculations. This value was determined based on the experimentally measured $T_c$ of 90 K for (Y,La)H$_4$ at 110 GPa from previous studies~\cite{BI2022100840}.

%To investigate $T_c$ for (Y,Sc)H$_4$, (Y,Zr)H$_4$, and (Y,La)H$_4$, we selected $\mu^*$ = 0.05 based on the experimentally reported $T_c$ = 90 K for (Y,La)H$_4$ at 110 GPa in previous work~\cite{BI2022100840}.

%After the computational parameters were determined, the $T_c$ for these three materials were obtained to be 107.10 K, 62.67 K and 82.49 K, by the QE~\cite{Giannozzi_2009} code at 100 GPa, 200 GPa and 120 GPa. As seen in Fig.~\ref{fig5}(b), Fig.~\ref{fig5}(e) and Fig.~\ref{fig5}(h), the $T_c$ calculated by the ELK~\cite{elk} program follows the same trend as the QE code and decreases as $\mu^*$ increases. Assuming a calculation parameter of 0.10, the $T_c$ is 104.36 K for (Y,Sc)H$_4$, 52.61 K for (Y,Zr)H$_4$, and 82.87 K for (Y,La)H$_4$, respectively. We selected $\mu^*$ = 0.05 based on the experimentally reported $T_c$ = 90 K for (Y,La)H$_4$ at 110 GPa in previous work~\cite{BI2022100840}.

Once the computational parameters were established, $T_c$ values for the two materials were calculated using the EPW code~\cite{Giannozzi_2017} code to be 104.32 K, 77.40 K, and 85.39 K at 100 GPa, 200 GPa, and 120 GPa, respectively. As shown in Fig.~\ref{fig5}(b), (d) and (f), $T_c$ values obtained using the ELK ~\cite{elk} program exhibit the same decreasing trend with increasing $\mu^*$ as observed with the EPW code. Assuming $\mu^*$= 0.05, the calculated $T_c$ values are 124.43 K for (Y,Sc)H$_4$, 69.55 K for (Y,Zr)H$_4$, and 101.24 K for (Y,La)H$_4$. The two computational approaches yield consistent results, indicating the credibility of our findings on the superconducting properties of these materials.

After systematically evaluating two computational approaches, we ultimately selected the ELK~\cite{elk} code results for final reporting due to their optimal balance between computational efficiency and predictive accuracy. While the superconducting transition temperature ($T_c$) of (Y,La)H$_4$ has been previously documented~\cite{BI2022100840}, our calculations predict that (Y,Sc)H$_4$ at 100 GPa exhibits a significantly enhanced $T_c$ of 124.43 K. This theoretical prediction, representing a notable advancement beyond currently reported values, awaits experimental verification through future high-pressure studies.

\begin{figure}
\includegraphics[width=1.0\columnwidth]{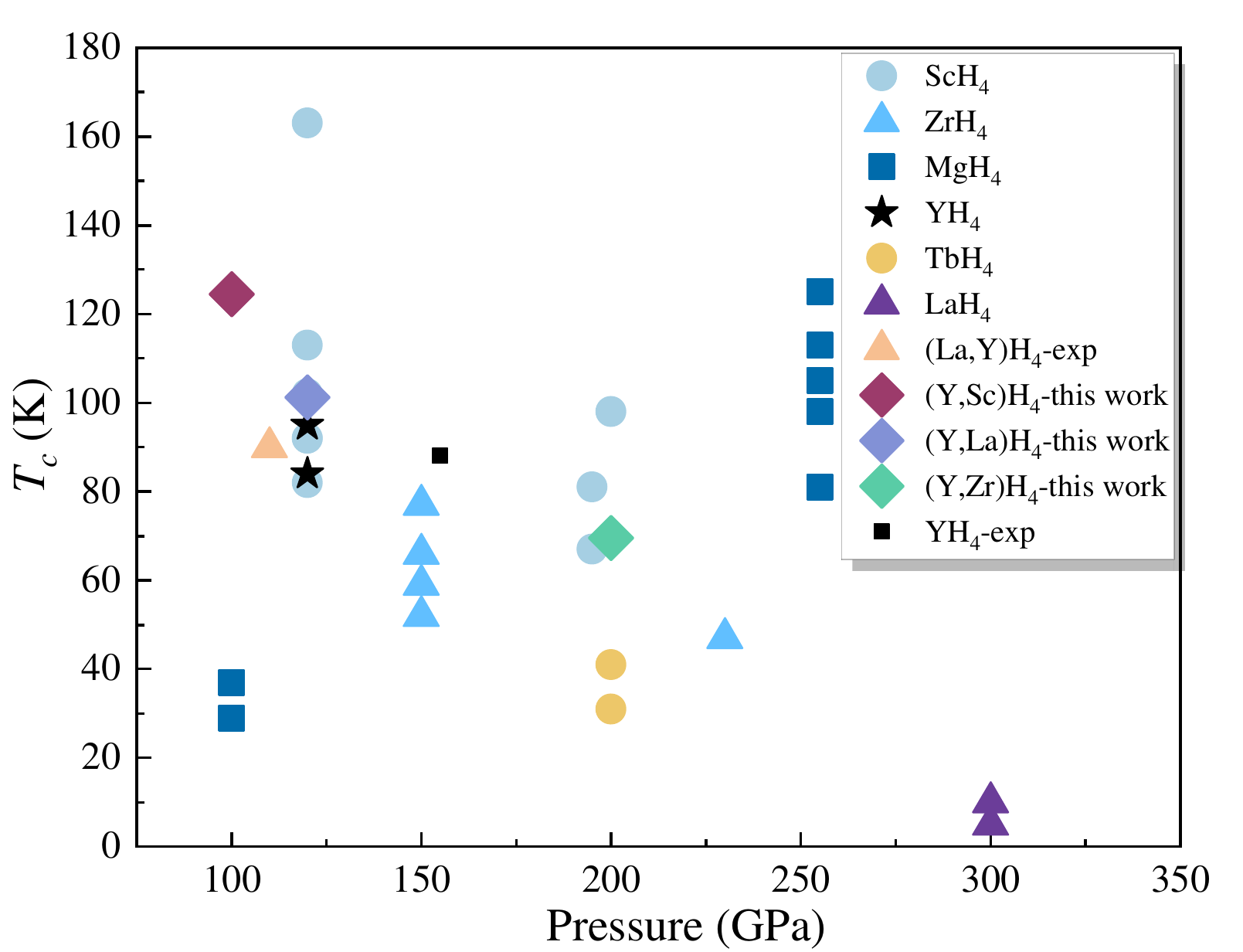}
%\includegraphics[width=\columnwidth]{fig6}
%\vspace{-2mm}
\caption{%
%The $T_c$ of MH$_4$ (M for different metal elements) the space group for $I$4/$mmm$ at different pressures, where the red, purple, and green rhombuses represent (Y,Sc)H$_4$, (Y,Zr)H$_4$ and (Y,La)H$_4$. Light blue dots, bright blue triangles, navy blue squares, black stars, orange dots, purple triangles, and pink triangles represent ScH$_4$~\cite{ye2018high,PhysRevB.96.144108,PhysRevB.96.094513,https://doi.org/10.1002/chem.202102679},
%ZrH$_4$~\cite{https://doi.org/10.1002/chem.202102679,PhysRevB.98.134103}, MgH$_4$~\cite{https://doi.org/10.1002/chem.202102679,PhysRevB.98.134103,PhysRevB.87.054107}, YH$_4$~\cite{li2015pressure}, TbH$_4$~\cite{doi:10.1021/acs.jpcc.1c00645}, LaH$_4$~\cite{doi:10.1073/pnas.1704505114} and (La,Y)H$_4$~\cite{BI2022100840}, respectively.
%
%represent ScH$_4$~\cite{ye2018high,PhysRevB.96.144108,PhysRevB.96.094513,https://doi.org/10.1002/chem.202102679}, light blue triangles represent ZrH$_4$~\cite{https://doi.org/10.1002/chem.202102679,PhysRevB.98.134103}, navy blue squares represent MgH$_4$~\cite{https://doi.org/10.1002/chem.202102679,PhysRevB.98.134103,PhysRevB.87.054107}, black stars represent YH$_4$~\cite{li2015pressure}, orange dots represent TbH$_4$~\cite{doi:10.1021/acs.jpcc.1c00645}, purple triangles represent LaH$_4$~\cite{doi:10.1073/pnas.1704505114},and pink triangles represent (La,Y)H$_4$~\cite{BI2022100840}.
%
Critical temperature $T_c$ of MH$_4$ compounds (M represents different metal elements) with the $I$4/$mmm$ space group at various pressures. The red, purple, and green rhombuses indicate $T_c$ values for (Y,Sc)H$_4$, (Y,Zr)H$_4$ and (Y,La)H$_4$, respectively. Light blue dots, bright blue triangles, navy blue squares, black stars, orange dots, purple triangles, and pink triangles represent data for ScH$_4$~\cite{doi:10.1021/acs.jpcc.7b12124,PhysRevB.96.144108,PhysRevB.96.094513,https://doi.org/10.1002/chem.202102679},
ZrH$_4$~\cite{https://doi.org/10.1002/chem.202102679,PhysRevB.98.134103}, MgH$_4$~\cite{https://doi.org/10.1002/chem.202102679,PhysRevB.98.134103,PhysRevB.87.054107}, YH$_4$~\cite{li2015pressure,PhysRevB.104.174509}, TbH$_4$~\cite{doi:10.1021/acs.jpcc.1c00645}, LaH$_4$~\cite{liu2017potential} and (La,Y)H$_4$~\cite{BI2022100840}, respectively.
\label{fig6}}
\end{figure}

To compare the $T_c$ of binary hydrides in the $I$4/$mmm$ space group under varying pressure conditions, we compiled their data alongside our results, as shown in Fig.~\ref{fig6}. In the field of superconductivity, the ideal materials exhibit higher $T_c$ under lower pressures. As illustrated in Fig.~\ref{fig6}, the $T_c$ for (Y,Sc)H$_4$ in this study, 124.43 K, establishes it as the best-performing superconductor among binary hydrides in the $I$4/$mmm$ space group. Our findings indicate that (Y,Sc,La)H$_4$ outperforms pure YH$_4$ ~\cite{li2015pressure,PhysRevB.104.174509}, ScH$_4$~\cite{doi:10.1021/acs.jpcc.7b12124,PhysRevB.96.144108,PhysRevB.96.094513,https://doi.org/10.1002/chem.202102679}, and LaH$_4$~\cite{liu2017potential}. Conversely, (Y,Zr)H$_4$ exhibits a significantly lower $T_c$ compared to other compounds in the same family. Based on these results, we conclude that metallic elements can be incorporated into binary hydride superconductors to enhance $T_c$ and reduce the required pressure. A particularly effective strategy involves substituting a metal element from the same group, while substitution with neighboring elements significantly degrades the superconducting performance of materials.

%===========< FIGURE 6 >=========================================
%\begin{figure*}
%\end{figure*}
%===========< FIGURE 6 >=========================================

\section{IV. CONCLUSIONS AND DISCUSSIONS}
%In this work, we determine the precise superconductivity $T_c$ for (Y,Sc)H$_4$ (100 GPa), (Y,Zr)H$_4$ (200 GPa) and (Y,La)H$_4$ (120 GPa) by substituting Sc, Zr, and La for half of the Y elements using three different codes. Additionally, we have calculated electronic band structures, phonon dispersions, electron-phonon coupling strengths, and superconducting property based on the McMillan-Allen-Dynes formula, the isotropic Migdal-Eliashberg and the anisotropic Migdal-Eliashberg equations for (Y,Sc)H$_4$, (Y,Zr)H$_4$ and (Y,La)H$_4$. We have obtained the following key findings.

In this work, we determined the precise superconducting critical temperatures $T_c$ for (Y,Sc)H$_4$ (100 GPa), (Y,Zr)H$_4$ (200 GPa) and (Y,La)H$_4$ (120 GPa) by substituting Sc, Zr, and La for half of the Y atoms, using three computational codes. Additionally, we calculated electronic band structures, phonon dispersions, electron-phonon coupling strengths, and superconducting properties based on the McMillan-Allen-Dynes formula, as well as the isotropic and anisotropic Migdal-Eliashberg equations. Our key findings are summarized as follows:

%(i)	For (Y,Sc)H$_4$ (100 GPa), (Y,Zr)H$_4$ (200 GPa), and (Y,La)H$_4$ (120 GPa), the $T_c$ are 104.36 $\sim$ 143 K, 52.61 $\sim$ 100 K, and 82.49 $\sim$ 128 K when $\mu^*$ = 0.1, respectively. 
(i) For (Y,Sc)H$_4$ (100 GPa), (Y,Zr)H$_4$ (200 GPa), and (Y,La)H$_4$ (120 GPa), the $T_c$ values are 124.43 K, 69.55 K, and 101.24 K, respectively, when $\mu^*$ = 0.05.

%Furthermore, as $\mu^*$ falls, it is predicted that the size of $T_c$ will rise.

%(ii) We trace the excellent superconductivity of (Y, Sc, La, Zr)H$_4$ to the crucial contribution of lowest phonon branches.
(ii) The excellent superconductivity of (Y,Sc,Zr,La)H$_4$ can be attributed to the critical contribution of the lowest phonon branches.

%(iii) On the one hand, (Y,Sc)H$_4$ has no gap between the optical and acoustic phonon branches. On the other hand, it still maintains a bond length at 100 GPa similar to that of (Y,La)H$_4$ at 120 GPa and (Y,Zr)H$_4$ at 200 GPa. All these indicate that (Y,Sc)H$_4$ has a significant electron-phonon coupling and good superconductivity.
%(iii) (Y,Sc)H$_4$ stands out due to the absence of a gap between its optical and acoustic phonon branches, while maintaining a bond length at 100 GPa comparable to that of (Y,La)H$_4$ at 120 GPa and (Y,Zr)H$_4$ at 200 GPa. These factors contribute to its significant electron-phonon coupling and superior superconducting properties.
(iii) The strong electron-phonon coupling in (Y,Sc)H$_4$ arises from a combination of soft hydrogen-dominated phonon modes, strong covalent bonding between hydrogen and metal atoms, and electron localization in interstitial regions.

%(iv) The selection of alternative metal elements with the same group for doping can further improve the superconducting temperature, while the neighboring elements significantly degrades the superconductivity of the material.
%(iv) Substituting alternative metal elements within the same group for doping effectively enhances the superconducting critical temperatures $T_c$, whereas doping with neighboring elements significantly degrades the superconducting performance of material.
(iv) Doping with isovalent metals enhances $T_c$, while adjacent-element doping degrades superconducting performance.

%The unique superconducting characteristics of (Y,Sc)H$_4$ (100 GPa) and (Y,La)H$_4$ (120 GPa), like those of YH$_4$, may be explained by the fact that their doping elements belong to the same subgroup as Y. This indicates that the hydrides that Y forms with these two dopant elements are more stable, but the connection with Y and Zr does not work out as well as anticipated. According to the EDOS value at the Fermi level, the greater fraction of H-dominated frequency phonons and the shorter bond lengths of the Y and Sc bonds may be the causes of the high $T_c$ of (Y,Sc)H$_4$.

\section{SUPPLEMENTARYMATERIAL}
See the supplementary material for unit cell construction in Sc-, La-, and Zr-substituted systems, pressure selection criteria, and dynamic and thermodynamic stability analysis of (Y,Sc)H$_4$, (Y,Zr)H$_4$, and (Y,La)H$_4$. The Electron localization function (ELF) plots at corresponding pressures are also presented.

\section{ACKNOWLEDGMENTS}
The authors gratefully acknowledge discussions with Zhen Tong about the EPW code
and Wenbo Zhao about the ELK code.
We acknowledge the support from the National Natural Science Foundation of China 
(No.12104356 and No.52250191). % Zhibin Gao
This work is sponsored by the Key Research and Development Program of the Ministry of Science and Technology
(No.2023YFB4604100).
We also acknowledge the support by HPC Platform, Xi’an Jiaotong University.

%+++++++++++++++++++++++++++++++++++++++++++++++++++++++++++++++++++++
% You should use BibTeX and apsrev.bst for references
% Choosing a journal automatically selects the correct APS
% BibTeX style file (bst file), so only uncomment the line
% below if necessary.
% \bibliographystyle{apsrev4-1}
 \bibliography{YScH4} %your bib file here
 \end{document}